\documentclass[9pt,twocolumn,twoside]{opticajnl}

\journal{josab} 

\setboolean{shortarticle}{false}

\usepackage{amssymb,amsmath}
\usepackage{bm}
\usepackage{threeparttablex}
\usepackage{tabularx}
\usepackage{multirow}
\usepackage{siunitx}
\usepackage[justification=justified]{caption}

\usepackage{tikz,xcolor,hyperref}
\definecolor{lime}{HTML}{A6CE39}  
\DeclareRobustCommand{\orcidicon}{%
	\begin{tikzpicture}
	\draw[lime, fill=lime] (0,0) 
	circle [radius=0.16] 
	node[white] {{\fontfamily{qag}\selectfont \tiny ID}};
	\draw[white, fill=white] (-0.0625,0.095) 
	circle [radius=0.007];
	\end{tikzpicture}
	\hspace{-2mm}
}
\foreach \x in {A, ..., Z}{%
	\expandafter\xdef\csname orcid\x\endcsname{\noexpand\href{https://orcid.org/\csname orcidauthor\x\endcsname}{\noexpand\orcidicon}}
}

\renewcommand{\eqref}[1]{\hyperref[{#1}]{\textup{(\ref*{#1}})}}
\newcommand{\figref}[1]{\hyperref[{#1}]{\textup{Fig.~\ref*{#1}}}}
\newcommand{\secref}[1]{\hyperref[{#1}]{\textup{Sec.~\ref*{#1}}}}
\newcommand{\tabref}[1]{\hyperref[{#1}]{\textup{Table~\ref*{#1}}}}
\renewcommand{\algref}[1]{\hyperref[{#1}]{\textup{Algorithm~\ref*{#1}}}}

\newcommand{\vect}[1]{\mathbf{#1}}
\newcommand{\uvec}[1]{\widehat{\bm{#1}}}
\newcommand{\tensor}[1]{\overline{\overline{#1}}}
\newcommand{\im}{\text{i}}

\allowdisplaybreaks  

\title{Analysis and design of transition radiation in layered uniaxial crystals using Tandem neural networks}

\author[1]{Xiaoke Gao\orcidA}
\author[1]{Xiaoyu Zhao\orcidB}
\author[1]{Ruoyu Huang\orcidC}
\author[1]{Siyuan Ma\orcidD}
\author[1]{Xikui Ma}
\author[1,*]{\\Tianyu Dong\orcidF} 

\affil{School of Electrical Engineering, Xi'an Jiaotong University, Xi'an 710049, China}
\affil[*]{Corresponding author: tydong@mail.xjtu.edu.cn}

\begin{abstract}
With the flourishing development of nanophotonics, Cherenkov radiation pattern can be designed to achieve superior performance in particle detection by fine-tuning the properties of metamaterials such as photonic crystals (PCs) surrounding the swift particle. However, the radiation pattern can be sensitive to the geometry and material properties of PCs, such as periodicity, unit thickness, and dielectric fraction, making direct analysis and inverse design difficult. In this article, we propose a systematic method to analyze and design PC-based transition radiation, which is assisted by deep learning neural networks. By matching boundary conditions at the interfaces, Cherenkov-like radiation of multilayered structures can be resolved analytically using the cascading scattering matrix method, despite the optical axes not being aligned with the swift electron trajectory. Once well trained, forward deep learning neural networks can be utilized to predict the radiation pattern without further direct electromagnetic simulations; moreover, Tandem neural networks have been proposed to inversely design the geometry and/or material properties for desired Cherenkov radiation pattern. Our proposal demonstrates a promising strategy for dealing with layered-medium-based Cherenkov radiation detectors, and it can be extended for other emerging metamaterials, such as photonic time crystals.
\end{abstract}
\setboolean{displaycopyright}{true}
\doi{\url{http://dx.doi.org/10.1364/josab.XXXXXX}}
\dates{Compiled November 24, 2022}

\begin{document}
\maketitle

\section{Introduction}
Since its discovery in the 1930's, Cherenkov radiation has demonstrated a variety of applications, such as the detection of high-energy charged particles \cite{jelley1955cerenkov} and Cherenkov free-electron laser \cite{walsh1992radio}, labeled bio-molecules detection \cite{liu2010radiation}, molecular imaging \cite{glaser2015cherenkov} and cosmic rays detection \cite{icecube2013evidence}, to name a few. Recent applications of Cherenkov radiation in the emerging field of nanophotonics including surface plasmon excitation \cite{liu2012surface,yao2012thz,zhang2016tunable,zhong2018surface,su2019manipulating} and charged particle velocity detection \cite{bache2010optical,lin2018controlling,hu2022surface,gunay2020continuously}. Significant research efforts are concentrated on regulating the radiation behaviors and lowing or even eliminating the Cherenkov threshold through the use of metamaterials and metasurface, which have fascinating applications, \emph{e.g.}, hyperbolic-metamaterial-based intense Cherenkov radiation emitted by low-energy electrons \cite{fernandes2012cherenkov,liu2017integrated}, controlling the transition radiation angles with high sensitivity by photonic crystals (PCs) \cite{luo2003cerenkov,lin2018controlling}, controlling the Cherenkov radiation by the metasurface with one-dimensional sub-wavelength rotated apertures \cite{wintz2017anisotropic}, \emph{etc}. In addition, particle identification sensitivity can be increased by introducing material anisotropy for Cherenkov radiation \cite{ginis2014controlling} and surface Dyakonov–Cherenkov radiation \cite{hu2022surface}.

Recently, it has been demonstrated that the angle of Cherenkov radiation can be controlled with resonance transition radiation by modifying the photonic crystal band structure \cite{lin2018controlling}; additionally, the velocity direction of the swift charge can be distinguished by introducing uniaxial media \cite{hu2022surface}. Theoretically, Cherenkov radiation can be controlled by modifying the properties of multilayered (meta)materials surrounding the swift electron, such as the periodicity, thickness of the unit cells, dielectric permittivity, and fractions or filling ratios. In order to analyze and design Cherenkov radiation, electromagnetic simulation of such a structure is essential. Other than widely-used general numerical methods such as finite element methods or finite difference time domain methods in nanophotonics, analytical solution may be obtained for layered structures \cite{lin2018controlling}, even when the material is spatially dispersive \cite{benedicto2015numerical,dong2016optical,dong2020numerical}. Yet most work focus either on the plane wave interaction with layered structure comprised of anisotropic media \cite{yeh1980optics} or on the interaction of swift electron on 1D PCs comprised of isotropic medium \cite{lin2018controlling}, we have extended the powerful scattering matrix method (SMM) for simulating the Cherenkov transition radiation of multilayered structures that may contain anisotropic uniaxial crystals, even when the optical axis is not aligned to the electron trajectory.This numerical Swiss army knife for the Cherenkov radiation of multilayers provides not only a potent electromagnetic simulation tool, but also an easy-to-use and highly efficient data generator for the deep neural networks that will be presented in the following section.
 
Despite the fact that multilayered PC-based Cherenkov detectors have high performance when the transition resonances exhibit \cite{lin2018controlling}, it can be allergic to the material permittivity and the filling ratio. When a PC reduces to an ordinary multilayered structure due to fabrication error or the periodicity of PC changes, the above-mentioned direct analytical solution can be time-consuming with the large number of layers, which is unacceptable for the inverse design of such a multilayered structure. In this regard, the rapidly developing data-driven methods based on machine learning \cite{jordan2015machine,ray2019quick,carleo2019machine,ma2021analytical} offer promising solutions to such nanophotonic problems for both the direct simulation and the inverse design \cite{molesky2018inverse,ma2021deep,wiecha2021deep}. Rather than directly solving a problem for given parameters using conventional methods, one can generate a neural network (NN) whose training data are available according to present analytical or numerical methods, and new problems can be resolved by the well-trained deep neural networks even if the provided parameters do not appear in the training dataset. The NN-assisted methods have been successful in predicting the optical properties of a complex physical system, such as the extinction spectrum of multilayered nano-particles and meta-structures \cite{peurifoy2018nanophotonic,liu2018training} and the dispersion relation of metamaterials, \emph{etc.}; they are also capable of inverse design when the desired properties are required. Typically, an inverse problem presents multiple solutions for a specific desired response, which traditional neural networks cannot effectively solve because they are difficult to converge \cite{jiang2021deep}. The deterministic and generative modes , such as Tandem networks, variational auto-encoders (VAEs), and generative adversarial networks (GAN), \emph{etc}, have been introduced to the nanophotonic community in order to solve the convergence problems for inverse problems \cite{taigao2022benchmarking}. In this paper, a Tandem network containing the pre-trained simulation forward sub-net is proposed, which demonstrates excellent performance in the design of multilayer-PC-based meta-structures for effective Cherenkov radiation. 

Beginning with the transition radiation at a single interface \cite{ginzburg1979several,ginzburg1982transition}, we derive the boundary conditions for general anisotropic uniaxial crystals. Next, the generalized scattering matrix method that accounts for inhomogeneous source term has been proposed to calculate the interaction between a swift electron and a multilayered structure. In addition, by varying the angle between the electron trajectory and the optical axis of uniaxial crystals, we demonstrate the existence of a strong directional effective Cherenkov radiation in the plane perpendicular to the charge velocity. A Tandem network has also been proposed to manage the inverse design of PC-based effective Cherenkov-like radiation.

\section{Theory and Methods}
\begin{figure}[!ht]
    \centering
    \includegraphics{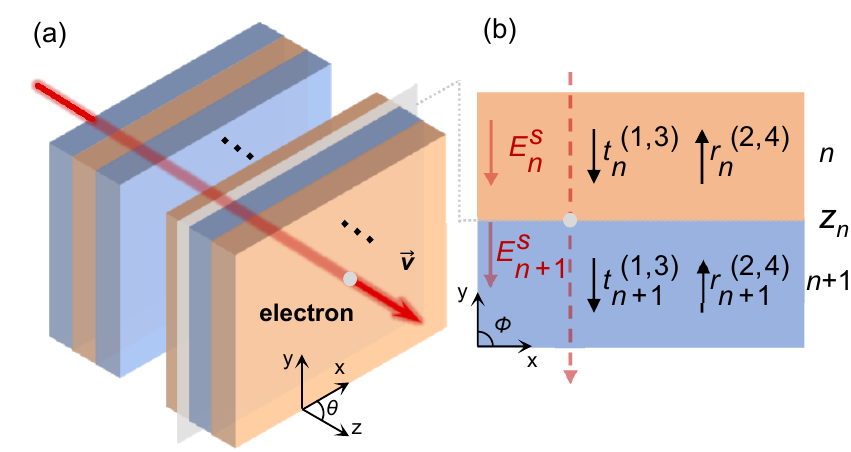}
    \caption{(a) Swift electrons traverse a multilayered structure; (b) Swift electrons moving from medium $n$ to medium $n+1$, where the planar interface is located at $z =z_n$.}
    \label{fig:figure01}
\end{figure}

\figref{fig:figure01} depicts the schematic of a swift electron traversing perpendicularly a single planar interface located at $z=z_n$, which is separated by two mediums denoted by $n$ and $n+1$. The interface normal is parallel to the electron's trajectory, \emph{i.e.}, $z$-axis. In general, $\tensor{\varepsilon}_d=\text{diag}(\varepsilon_1,\varepsilon_2,\varepsilon_3)$ represents the relative permittivity of the nonmagnetic medium, which can be isotropic, uniaxial or bi-axial. When the medium is isotropic, $\varepsilon_1=\varepsilon_2=\varepsilon_3$ and when it is uniaxial, $\varepsilon_1=\varepsilon_2\neq\varepsilon_3$. For uniaxial crystals, the optic axis may be deviated from the $z$-axis; therefore, the corresponding dielectric tensor reads $\tensor{\varepsilon} = \Lambda^{-1}\tensor{\varepsilon}_d\Lambda$, where $\Lambda = \Lambda(\alpha,\beta,\gamma)$ is the \emph{so-called} Euler rotation matrix ($y$-$z$-$y$ type) with $\alpha$, $\beta$ and $\gamma$ being the Euler angles (pitch, yaw, and roll angles, respectively) of the global coordinate axes and the ``optical axes'' determined by the uniaxial permittivity $\tensor{\varepsilon}_d$.

\subsection{Governing equations and the solutions}
Assuming that the velocity of the electron $\vect{v} = \uvec{z} v$ is unaffected by the radiation, the swift electron can be regarded as a current source which can be expressed as  $\vect{J}_e(\vect{r},t) = \hat{z}qv\delta(x)\delta(y)\delta(z-v t)$. Thus, the governing partial differential equation (PDE) for the electric field $\vect{E}_n(\vect{r},t)$ inside the $n$-th layer reads as
\begin{equation}\label{eq:MaxwellWaveEquation}
    \nabla\times\nabla\times\vect{E}_n(\vect{r},t) + \mu_0\varepsilon_0 \tensor{\varepsilon}_n \cdot \partial_t^2 \vect{E}_n(\vect{r},t) = -\mu_0 \partial_{t} \vect{J}_e(\vect{r},t),
\end{equation}
where $\mu_0$ and $\varepsilon_0$ are the permeability and permittivity of the free space, respectively; $\tensor{\varepsilon}_n$ is the relative dielectric tensor of the $n$-th layer;  $\partial_t$ and $\partial_t^2$ denote the first- and second-order time derivatives, respectively. 

For the considered multilayered structure, instead of handling the wave equation \eqref{eq:MaxwellWaveEquation} in the spatial-temporal domain, we can apply the Fourier transform defined as
\begin{subequations}
\begin{equation} \label{eq:FourierTransform}
    \vect{F}(k_x,k_y,z,\omega) = \int  \vect{F}(\vect{r},t) e^{\im (\omega t - k_x x - k_y y)} \text{d}t \text{d}x \text{d}y
\end{equation}
to the source $\vect{J}_e(\vect{r},t)$ and the fields $\vect{E}_n(\vect{r},t)$, with the inverse Fourier transform reading as 
\begin{equation} \label{eq:inverseFourierTransform}
    \vect{F}(\vect{r},t) = \left(\frac{1}{2\pi}\right)^{3} \int \vect{F}(k_x,k_y,z,\omega) e^{-\im (\omega t - k_x x - k_y y)} \text{d}\omega \text{d}k_x \text{d}k_y.
\end{equation}
\end{subequations}
Here, an $e^{-\im \omega t}$ time harmonic is adopted throughout. Note that, although the same kernels $\vect{F}$ are shared in \eqref{eq:FourierTransform} and \eqref{eq:inverseFourierTransform}, they are essential different physical quantities for different domain of interest; and the arguments $(k_x,k_y,z,\omega)$ or $(\vect{r}, t)$ are typically omitted without introducing ambiguity hereafter. Therefore, the wave equation \eqref{eq:MaxwellWaveEquation} in the spatial-temporal domain can be rewritten in the hybrid domain $\{k_x,k_y,z,\omega\}$ as
\begin{equation} \label{eq:MaxwellWaveEquationInHybridDomain}
    \begin{pmatrix}
    k_y^2 - \frac{\text{d}^2}{\text{d}z^2} & -k_x k_y & \im k_x \cdot \frac{\text{d}}{\text{d}z} \\
    -k_y k_x & k_x^2 - \frac{\text{d}^2}{\text{d}z^2} & \im k_y \cdot \frac{\text{d}}{\text{d}z} \\
    \frac{\text{d}}{\text{d}z} \cdot \im k_x  & \frac{\text{d}}{\text{d}z} \cdot \im k_y & k_x^2 + k_y^2
    \end{pmatrix} \cdot
    \vect{E}_n - \tensor{\varepsilon}_n k_0^2 \cdot \vect{E}_n = \im\omega\mu_0 \vect{J}_e,
\end{equation}
where $\vect{J}_e(k_x,k_y,z,\omega) = \hat{z} q e^{\im k_e z}$ with $k_e = \omega/v$; $k_0^2=\omega^2\mu_0\epsilon_0$ denotes the wave number in free space; $\text{d}/\text{d}z$ and $\text{d}^2/\text{d}z^2$ denote the first- and second-order spatial derivatives with respect to the variable $z$, respectively. Now, the ordinary differential equations (ODEs)  \eqref{eq:MaxwellWaveEquationInHybridDomain} can be solved in the hybrid domain for different wavenumbers $k_x$ and $k_y$ at different frequencies $\omega$, and the final solutions to the original PDEs \eqref{eq:MaxwellWaveEquation} can be readily obtained using the inverse Fourier transform \eqref{eq:inverseFourierTransform}. For the inhomogeneous ODE \eqref{eq:MaxwellWaveEquationInHybridDomain}, the solution consists of the so-called the general solutions $\vect{E}_{n,g}$ and the particular solution $\vect{E}_{n,p}$, which correspond to the transverse electromagnetic modes and the longitudinal electromagnetic mode excited by the moving charge, respectively.

\paragraph{General solution.}
For the general solutions $\vect{E}_{n,g}$ to \eqref{eq:MaxwellWaveEquationInHybridDomain}, they satisfy $\text{d}\vect{E}_{n,g}/\text{d}z = \im k_{z,n}\vect{E}_{n,g}$. As a result, a set of linear algebraic equations can be derived, which reads
\begin{equation}
    \mathbb{M}_{n,g}(k_x,k_y,k_{z,n},k_0)\cdot\vect{E}_{n,g} = 0,
\end{equation}
where
\begin{equation} \label{eq:MatrixMg}
    \mathbb{M}_{n,g}(k_x,k_y,k_{z,n},k_0) = \tensor{\varepsilon}_n k_0^2 -
    \begin{pmatrix}
        k_y^2+k_{z,n}^2 & -k_x k_y & -k_x k_{z,n} \\
        -k_y k_x & k_x^2+k_{z,n}^2 & -k_y k_{z,n} \\
        -k_{z,n} k_x & -k_{z,n} k_y & k_x^2+k_y^2
    \end{pmatrix} .
\end{equation}
For any non-trivial modes $\vect{E}_{n,g}$, the zero determinant of the matrix $\mathbb{M}_{n,g}(k_x,k_y,k_{z,n},k_0)$ yields the dispersion relation of the material, \emph{i.e.},
\begin{equation} \label{eq:dispersionRelation}
    \text{det}[\mathbb{M}_{n,g}(k_x,k_y,k_{z,n},k_0)] = 0,
\end{equation}
which is a quartic polynomial equation with respect to $k_z$, whose solutions are denoted by $k_{z,n}^{(i)}$ ($i=1, 2, 3, 4$), indicating that four different waves could exhibit. Accordingly, the corresponding polarization modes $\vect{E}_{n}^{(i)}$ can be readily obtained by evaluating the null space of the matrix $\mathbb{M}_{n,g}^{(i)}(k_x,k_y,k_{z,n}^{(i)},k_0)$, \emph{i.e.},
\begin{equation} \label{eq:Modes}
    \vect{E}_{n}^{(i)} = \text{NullVector}[\mathbb{M}_{n,g}(k_x,k_y,k_{z,n}^{(i)},k_0)],
\end{equation}
which corresponds to the wave vector $\vect{k} = (k_x,k_y,k_{z,n}^{(i)})$. Specifically, when the material is isotropic, modes will be degenerated. In the subsequent discussion, we assume that $k_{z,n}^{(1)}$ ( $k_{z,n}^{(3)}$) corresponds the forward wave and $k_{z,n}^{(2)}$ ($k_{z,n}^{(4)}$) corresponds the backward wave. As a result, the general solutions to \eqref{eq:MaxwellWaveEquationInHybridDomain} can be expressed in terms of the independent modes $\vect{E}_{n}^{(i)}$.

\paragraph{Particular solution.}
For the particular solution $\vect{E}_{n,s}$ to \eqref{eq:MaxwellWaveEquationInHybridDomain}, it satisfies $\text{d}\vect{E}_{n,s}/\text{d}z = \im k_e\vect{E}_{n,s}$ since the inhomogeneous source term $-\im\omega\mu_0\vect{J}_e = -\uvec{z} \im\omega\mu_0 q e^{\im k_e z}$ is a harmonic function with respect to $z$. Thus, one can obtain a set of linear algebraic equations from \eqref{eq:MaxwellWaveEquationInHybridDomain}, which reads as
\begin{equation}
    \mathbb{M}_{n,s}(k_x,k_y,k_0)\cdot\vect{E}_{n,s} = -\im\omega\mu_0\vect{J}_e,
\end{equation}
where $\mathbb{M}_{n,s}(k_x,k_y,k_0) = \mathbb{M}_{n,g}(k_x,k_y,k_e, k_0)$. As a result, the particular solution to \eqref{eq:MaxwellWaveEquationInHybridDomain} can be directly obtained as
\begin{equation} \label{eq:SolutionEp}
    \vect{E}_n^s = -\im\omega\mu_0\mathbb{M}_{n,s}^{-1}(k_x,k_y,k_0)\cdot\vect{J}_e.
\end{equation}

\subsection{Boundary conditions} 
Given the general solution $\vect{E}_{n,g}$ for the transverse wave and the particular solution $\vect{E}_n^s$ for the longitudinal wave, the overall electric field in each layer  $\vect{E}_n(\vect{r})$ can be expressed in terms of theses modes $\vect{E}_{n,g}$ and $\vect{E}_n^s$, which reads as
\begin{equation}\label{eq:ElectricFieldExpression}
\begin{split}
    \vect{E}_n(z) = & t^{(1)}_n\vect{E}_n^{(1)} e^{\im k_{z,n}^{(1)} z} + r^{(2)}_n\vect{E}_n^{(2)} e^{\im k_{z,n}^{(2)} z} \\ 
    &+ t^{(3)}_n\vect{E}_n^{(3)} e^{\im k_{z,n}^{(3)} z} + r^{(4)}_n\vect{E}_n^{(4)} e^{\im k_{z,n}^{(4)} z} + \vect{E}_n^s e^{\im k_e z},
\end{split}
\end{equation}
where $t^{(1)}_n$ ($t^{(3)}_n$) denotes the amplitude of the forward wave $\vect{E}_n^{(1)}$ ($\vect{E}_n^{(3)}$), and $r^{(2)}_n$ ($r^{(4)}_n$) denotes the amplitude of the backward wave $\vect{E}_n^{(2)}$ ($\vect{E}_n^{(4)}$). Accordingly, the magnetic field can be expressed as  
\begin{equation}\label{eq:MagneticFieldExpression}
\begin{split}
    \vect{H}_n(z) = & t^{(1)}_n\vect{H}_n^{(1)} e^{\im k_{z,n}^{(1)} z} + r^{(2)}_n\vect{H}_n^{(2)} e^{\im k_{z,n}^{(2)} z} \\
    &+ t^{(3)}_n\vect{H}_n^{(3)} e^{\im k_{z,n}^{(3)} z}+ r^{(4)}_n\vect{H}_n^{(4)} e^{\im k_{z,n}^{(4)} z} + \vect{H}_n^s e^{\im k_e z},
\end{split}
\end{equation}
where $\vect{H}_n^{(i)} = (\omega\mu_0)^{-1} (k_x,k_y,k_{z,n}^{(i)})^\text{T} \times \vect{E}_n^{(i)}$ and $\vect{H}_n^s = (\omega\mu_0)^{-1} (k_x,k_y,k_e)^\text{T} \times \vect{E}_n^s$. Yet unknown, these mode amplitudes $t$'s and $r$'s can be resolved by matching boundary conditions at each interface, which will be discussed next.

As shown in \figref{fig:figure01}(b), at the planar interface located at $z=z_n$ which separates the media $\tensor{\varepsilon}_n$ and $\tensor{\varepsilon}_{n+1}$, the tangential components of  $\vect{E}_n$ and $\vect{H}_n$ are respectively continuous, yielding a set of equations that can be expressed in a compact matrix form with respect to the mode coefficients $t$'s and $r$'s as
\begin{equation} \label{eq:boundaryConditionsAtSingleInterface}
    \mathcal{D}_{n,n+1} \cdot \mathcal{P}_{n,n+1} \cdot
                        \mathcal{C}_{n,n+1} 
                          + \bm{f}_n = 
    \mathcal{D}_{n+1,n} \cdot \mathcal{P}_{n+1,n} \cdot
                         \mathcal{C}_{n+1,n}
                          + \bm{f}_{n+1}.
\end{equation}
where
\begin{subequations}\label{eq:boundaryConditionAtSingleInterfaceAddition}
\begin{align}
    \mathcal{D}_{l,m} &=
    \begin{pmatrix}
      \uvec{x}\cdot \vect{E}_l^{(1)}  & \uvec{x}\cdot \vect{E}_l^{(3)} & -\uvec{x}\cdot \vect{E}_{m}^{(2)} & -\uvec{x}\cdot \vect{E}_{m}^{(4)}\\
      \uvec{y}\cdot \vect{E}_l^{(1)} & \uvec{y}\cdot \vect{E}_l^{(3)} & -\uvec{y}\cdot \vect{E}_{m}^{(2)} & -\uvec{y}\cdot \vect{E}_{m}^{(4)}\\
      \uvec{x}\cdot \vect{H}_l^{(1)} & \uvec{x}\cdot \vect{H}_l^{(3)} & -\uvec{x}\cdot \vect{H}_{m}^{(2)} & -\uvec{x}\cdot \vect{H}_{m}^{(4)}\\
      \uvec{y}\cdot \vect{H}_l^{(1)} & \uvec{y}\cdot \vect{H}_l^{(3)} & -\uvec{y}\cdot \vect{H}_{m}^{(2)} & -\uvec{y}\cdot \vect{H}_{m}^{(4)}
    \end{pmatrix}, \\
    \mathcal{P}_{l,m} &= \text{diag}(e^{\im k_{z,l}^{(1)}z_n}, \quad e^{\im k_{z,l}^{(3)}z_n}, \quad e^{\im k_{z,m}^{(2)}z_n}, \quad e^{\im k_{z,m}^{(4)}z_n}),\\
    \mathcal{C}_{l,m} &= \left( t_l^{(1)}, t_l^{(3)}, r_{m}^{(2)}, r_{m}^{(4)}\right)^{\mathrm{T}},
\end{align}
and
\begin{equation}
    \bm{f}_l = \left(\uvec{x}\cdot\vect{E}_l^s, \uvec{y}\cdot\vect{E}_l^s, \uvec{x}\cdot\vect{H}_l^s, \uvec{y}\cdot\vect{H}_l^s \right)^\mathrm{T} e^{\im k_e z_n}
\end{equation}
\end{subequations}
with $l$ and $m$ being $n$ or $n+1$; $\uvec{x}$ and $\uvec{y}$ being the unit vectors along the $x$- and $y$-directions, respectively. Evidently, the additional longitudinal wave caused by the swift electrons serving as the source enriches the transmission/reflection phenomena in such a structure with multiple layers. Although it is now possible to solve for the mode coefficients $t$'s and $r$'s in all layers, the linear system of equations grows large when the number of layers increases. To make large problems solvable, we intend to develop a matrix cascading technique.

\subsection{Generalized cascading scattering matrix method}
As described in literature \cite{yeh1980optics,benedicto2015numerical,dong2020numerical}, the transfer matrix method (TMM) and scattering matrix method (SMM) may be used for such multilayers under plane wave excitation, where the source terms $f$'s in \eqref{eq:boundaryConditionsAtSingleInterface} disappear for intermediate layers. Here, we have extended the SMM to account for the presence of $f$'s in an inhomogeneous scenario. 

We can rewrite the boundary equation \eqref{eq:boundaryConditionsAtSingleInterface} in the context of scattering matrix, which reads
\begin{equation}
    \mathcal{C}_{n+1,n} = \bm{S}_{n,n+1} \cdot \mathcal{C}_{n,n+1} + \bm{F}_{n,n+1},
\end{equation}
where the generalized scattering matrix $S_{n,n+1}$ and source matrix $F_{n,n+1}$ can be extracted from \eqref{eq:boundaryConditionsAtSingleInterface} :
\begin{subequations} \label{eq:ScatteringMatriesEstablish}
    \begin{align}
        \bm{S}_{n,n+1} &= \left( \mathcal{D}_{n+1,n}\cdot\mathcal{P}_{n+1,n} \right)^{-1}\cdot
        \left( \mathcal{D}_{n,n+1}\cdot\mathcal{P}_{n,n+1} \right),\\
        \bm{F}_{n,n+1} &=  \left( \mathcal{D}_{n+1,n}\cdot\mathcal{P}_{n+1,n} \right)^{-1}\cdot
         \left( \bm{f}_{n} - \bm{f}_{n+1} \right).
    \end{align}
\end{subequations}
Now, having two successive systems denoted by $\mathcal{C}_{m,l} = \bm{S}_{l,m}\cdot \mathcal{C}_{l,m} + \bm{F}_{l,m}$ and $\mathcal{C}_{n,m} = \bm{S}_{m,n}\cdot \mathcal{C}_{m,n} + \bm{F}_{m,n}$, one can readily cascade them and form a new system denoted by $\mathcal{C}_{n,l} = \bm{S}_{l,n}\cdot \mathcal{C}_{l,n} + \bm{F}_{l,n}$, where the elements of the cascading scattering matrix $\bm{S}_{l,n}$ and source vector $\bm{F}_{l,n}$ read as
\begin{subequations} \label{eq:CascadeFormula}
\begin{align}
    \bm{S}^{11}_{l,n} &= \bm{S}^{11}_{m,n} \cdot \chi_1 \cdot \bm{S}^{11}_{l,m},\\
    \bm{S}^{12}_{l,n} &= \bm{S}^{12}_{m,n} + \bm{S}^{11}_{m,n} \cdot \chi_1 \cdot \bm{S}^{12}_{l,m}\cdot \bm{S}^{22}_{m,n},\\
    \bm{S}^{21}_{l,n} &= \bm{S}^{21}_{l,m} + \bm{S}^{22}_{l,m} \cdot \chi_2 \cdot \bm{S}^{21}_{m,n}\cdot \bm{S}^{11}_{l,m},\\
    \bm{S}^{22}_{l,n} &= \bm{S}^{22}_{l,m} \cdot \chi_2 \cdot \bm{S}^{22}_{m,n},\\
    \bm{F}^1_{l,n} &= \bm{F}^1_{m,n} + \bm{S}^{11}_{m,n} \cdot \chi_1 \cdot (\bm{F}^1_{l,m} + S^{12}_{l,m} \cdot \bm{F}^2_{m,n} ),\\
    \bm{F}^2_{l,n} &= \bm{F}^2_{l,m} + \bm{S}^{22}_{l,m} \cdot \chi_2 \cdot (\bm{F}^2_{m,n} + S^{21}_{m,n} \cdot \bm{F}^1_{l,m} ).
\end{align}
\end{subequations}
Here, $\chi_1 = \left( \mathbb{I} - \bm{S}^{12}_{l,m} \cdot \bm{S}^{21}_{m,n} \right)^{-1}$ and $\chi_2 = \left( \mathbb{I} - \bm{S}^{21}_{m,n} \cdot \bm{S}^{12}_{l,m} \right)^{-1}$ with $\mathbb{I}$ denoting a $2\times2$ identity matrix; $\bm{S}^{ij}_{l,m}$ denotes the $(i,j)$-th $2\times2$ block matrix of the $4\times4$ $\bm{S}_{l,m}$; $\bm{F}^i_{l,m}$ denotes the $i$-th $2\times1$ sub-vector of $\bm{F}_{l,m}$. 

For a structure with multiple layers, we assume that the swift electron travels from layer 0 to layer $N$. Now, starting from $\bm{S}_{0,1}$, $\bm{S}_{1,2}$, $\bm{S}_{2,3}$, ..., $\bm{S}_{N-1,N}$ and $\bm{F}_{0,1}$, $\bm{F}_{1,2}$, $\bm{F}_{2,3}$, ..., $\bm{F}_{N-1,N}$, and by successively cascading the adjacent scattering matrices according to \eqref{eq:CascadeFormula}, we can obtain $\bm{S}_{0,2}$, $\bm{S}_{0,3}$, $\bm{S}_{0,4}$, ... and $\bm{F}_{0,2}$, $\bm{F}_{0,3}$, $\bm{F}_{0,4}$, ..., and finally the overall scattering matrix $\bm{S}_{0,N}$ and the inhomogeneous term $\bm{F}_{0,N}$. Thus, the overall backward reflection coefficients $\bm{r}_0^{(2,4)} = \left(r_0^{(2)}, r_0^{(4)} \right)^\text{T}$ and forward transmission coefficients $\bm{t}_N^{(1,3)} = \left(t_N^{(1)}, t_N^{(3)} \right)^\text{T}$ of the multilayered structure can readily be derived according to $\bm{S}_{0,N}$ and $\bm{F}_{0,N}$, which reads as
\begin{equation} \label{eq:overallTransmissionReflection}
    \begin{pmatrix} \bm{t}^{(1,3)}_{N} \\ \bm{r}^{(2,4)}_0 \end{pmatrix} =
    \begin{pmatrix}
        \bm{S}^{11}_{0,N} & \bm{S}^{12}_{0,N} \\
        \bm{S}^{21}_{0,N} & \bm{S}^{22}_{0,N}
    \end{pmatrix}
    \begin{pmatrix} \bm{t}^{(1,3)}_0 \\ \bm{r}^{(2,4)}_{N} \end{pmatrix}
    + \begin{pmatrix} \bm{F}^1_{0,N} \\ \bm{F}^2_{0,N} \end{pmatrix},
\end{equation}
where the transmission coefficients $\bm{t}_0^{(1,3)} = \left(t_0^{(1)}, t_0^{(3)} \right)^\text{T}$ in the entrance layer 0 are typically known and the reflection coefficients $\bm{r}_N^{(2,4)} = \left(r_N^{(2)}, r_N^{(4)} \right)^\text{T}$ in the exit layer $N$ are zero. \algref{alg:AnalyticalGCSMM} illustrates the corresponding pseudo code. In addition, it is simple to obtain the mode coefficients for the intermediate layers \cite{dong2016optical}. 
\begin{algorithm} 
    \caption{Calculation of radiation properties excited by the swift particle in multilayered structures.}\label{alg:AnalyticalGCSMM}
    \begin{algorithmic}[1]
    \Statex \textbf{Input:} Angles of interest: $\theta$, $\phi$; Frequency: $\omega$; Interface locations: $z_n$; Layer permittivity: $\tensor{\varepsilon}_n$; \# of interfaces: $N$. 
    \Statex \textbf{Output:} Forward/backward mode coefficients $\bm{t}_N^{(1,3)}, \bm{r}_0^{(2,4)}$.
    \For{$m \gets 1:N_m$}        \Comment{Compute modes; $N_m$: \# of media types }
        \State $\vect{E}_m^{(i)},\vect{H}_m^{(i)},\vect{k}_{m}^{(i)} \gets \theta, \phi, \omega, \tensor{\varepsilon}_m$  \Comment{With \eqref{eq:dispersionRelation}, \eqref{eq:Modes} and \eqref{eq:SolutionEp}. }
    \EndFor
    \For{$n \gets 1:N$}   \Comment{Form S-matrices at each interface}
        \State $S_{n-1,n},F_{n-1,n} \gets \vect{E}_{n-1}^{(i)},\vect{E}_n^{(i)},\vect{H}_{n-1}^{(i)},\vect{H}_n^{(i)},z_n$ \Comment{With \eqref{eq:ScatteringMatriesEstablish}.}
    \EndFor
    \For{$n \gets 2:N $}    \Comment{Cascade S-matrices}
        \State $S_{0,n},F_{0,n} \gets S_{0,n-1},F_{0,n-1}$ \Comment{With \eqref{eq:CascadeFormula}}
    \EndFor
    \State \textbf{return} $t_N^{(1)},t_N^{(3)},r_0^{(2)},r_0^{(4)}$    \Comment{With \eqref{eq:overallTransmissionReflection}}
    \end{algorithmic}
\end{algorithm}

\subsection{Angular spectral energy}
Now, the forward (backward) angular spectral energy density of PC-based Cherenkov-like transition radiation, defined as the radiation energy $W = \int_V \varepsilon E^2 \text{d}V$ per unit angular frequency $\omega$ per solid angle $\Omega$, reads $U(\theta,\phi,\omega) =\text{d}^{3}W/(\text{d}\omega\text{d}\theta\text{d}\phi)$. Here, the angles of interest $\theta$ and $\phi$ are related to the components of the ``wave vector'' $\vect{k}$ by $k_x = k_0 \sin{\theta}\cos{\phi}, k_y = k_0 \sin{\theta}\sin{\phi}$ where $k_0$ is the wave number of free space (entrance medium). Thus, the forward and backward spectral energy densities respectively read as \cite{lin2018controlling,hu2022surface}
\begin{subequations} \label{eq:AngularSpectralEnergy}
    \begin{align}
        U_{f}(\theta,\phi,\omega) &= \frac{1}{16\pi^3} \left|t^{(1)}_{N}\vect{E}_{N}^{(1)}+ t^{(3)}_{N}\vect{E}_{N}^{(3)} \right|^2
        \cdot\varepsilon_{N} k_{N}^2 \sin{2\theta},\\
        U_{b}(\theta,\phi,\omega) &= \frac{1}{16\pi^3} \left|r^{(2)}_{0}\vect{E}_0^{(2)}+ t^{(4)}_{0}\vect{E}_0^{(4)} \right|^2
        \cdot\varepsilon_0 k_0^2 \sin{2\theta},
    \end{align}
\end{subequations}
where $\varepsilon_0$ and $\varepsilon_{N}$ represent the permittivity in the initial (entrance) and final (exit) media, respectively; $k_0$ and $k_{N}$ denote the corresponding wave numbers. Using the elegant and powerful tool of generalized cascading scattering matrix method for swift electrons in multilayers, as illustrated in \algref{alg:AnalyticalGCSMM}, one can now evaluate the forward (backward) radiation in the bottom (top) free space region of the multilayers according to $U_{f(b)}(\theta,\phi,\omega)$. Consequently, unique Cherenkov radiation detectors can be designed by configuring the geometry and material properties of multilayers in a sophisticated manner. Thus, a Swiss army knife is desired for both analysis and design.

\subsection{Tandem-network-assisted analysis and design of multilayers}
If we view the problem of forward analysis as the process of evaluating $y=f(x)$, the inverse (design) problem $x=f^{-1}(y)$ is not straightforward as it may appear. Although optical responses such as transmission/reflection spectra and the Cherenkov-like radiation can be easily calculated for multilayer problems by analytical or numerical methods, the solution to the corresponding inverse problem may not exist. Such an inverse design in nanophotonics is typically a one-to-many problem, which makes it difficult to find a suitable solution $x$ for a given desired response $y$. Consequently, intelligent algorithms have been widely utilized to solve inverse problems.

For the majority of intelligent algorithms, including genetic algorithms, particle swarm and ant colony optimization techniques and neural networks (NNs), solving the inverse problem frequently requires repeatedly evaluating the forward problems with varying designing parameter values. When the forward problems' analysis methods are inefficient, the inverse problems become computationally expensive. Although the proposed generalized cascading scattering matrix method is more efficient than traditional numerical methods for multilayers, a deep neural network can be designed and trained to approximate the exact electromagnetic simulation (see \algref{alg:AnalyticalGCSMM}). Once the NN has been well trained, no further electromagnetic simulation is required; additionally, the NN is capable of predicting the electromagnetic response even if the designing parameters are not present in the training dataset, which is generated by using the proposed method (\algref{alg:AnalyticalGCSMM}). Consequently, we can use the more efficient NN to make the forward simulation solvable despite large number of layers. 

\begin{figure}[!ht]
    \centering
    \includegraphics{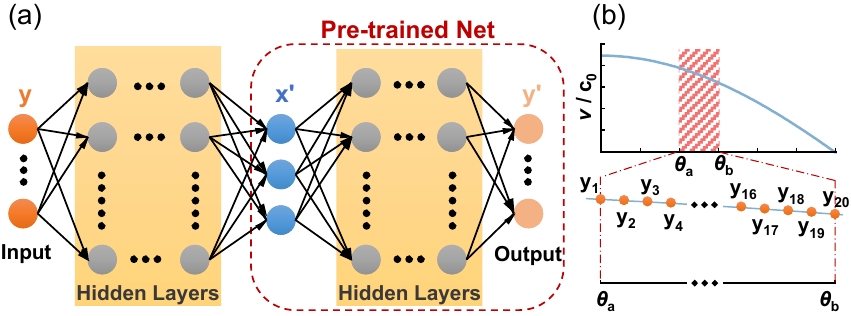}
    \caption{(a) Architecture of a Tandem network, which consists of two cascaded forward neural networks in serials. Given the expected optical response as the network input $y$, the designed response will be the output $y'$ which corresponds to the designed parameters $x'$ as the mid-products of the entire network. (b) Sampling method of the input sequence $y$, \emph{i.e.}, a portion of the detection curves of the Cherenkov-like radiation (see \emph{e.g.} \figref{fig:figure03}). }
    \label{fig:figure02}
\end{figure}

As for the corresponding inverse problem, one may consider adopting a similar structure of NN, with the desired electromagnetic properties as input and the designing parameters as output of the training dataset. Nevertheless, such a reverse NN frequently fails due to the one-to-many mapping nature of its inverse design. Although it is possible to improve the NN by carefully selecting the training dataset, this has little effect on the convergence performance for large training datasets \cite{liu2018training}. Consequently, variations of NN, such as Tandem networks\cite{liu2018training,ma2018deep}, variational auto-encoders (VAEs)\cite{sohn2015learning}, and generative adversarial networks (GANs) \cite{liu2018generative}, are regarded as effective solutions. \figref{fig:figure02}(a) illustrates a Tandem network composed of two cascaded fully-connected deep neural networks, one serving as the forward NN for the purpose of analysis and the other as the design NN. In constrast to the simple reverse NN, the error between the desired and designed electromagnetic responses, \emph{i.e.}, $|y-y'|$, is used for the back propagation algorithm during the training process, ensuring that the mid-products of the Tandem network $x'$ (the designed parameters) will produce a response that is nearly identical to the designed response as desired. In addition, the design object $x'$ in the Tandem network is no longer restricted to be as similar to the actual $x$ in the training dataset.

We shall mention that only a portion of parameter space for the angular response is used to train the Tandem network, whereas the designed parameters can produce a spectrum that closely matches the desired one across the whole entire range of angles of interest. This is a result of the physics-informed nature of the designed NN. During the training processing of the Tandem network, only the weights of the network for the inverse design will be updated, while the pre-trained forward network will remain unchanged. Although the proposed analytical model can be used as the forward model in the Tandem network, we retain the pre-trained NN because it is computationally inexpensive.

\section{Results and Discussion}
\subsection{Benchmark validation}
\begin{figure}[!ht]
    \centering
    \includegraphics{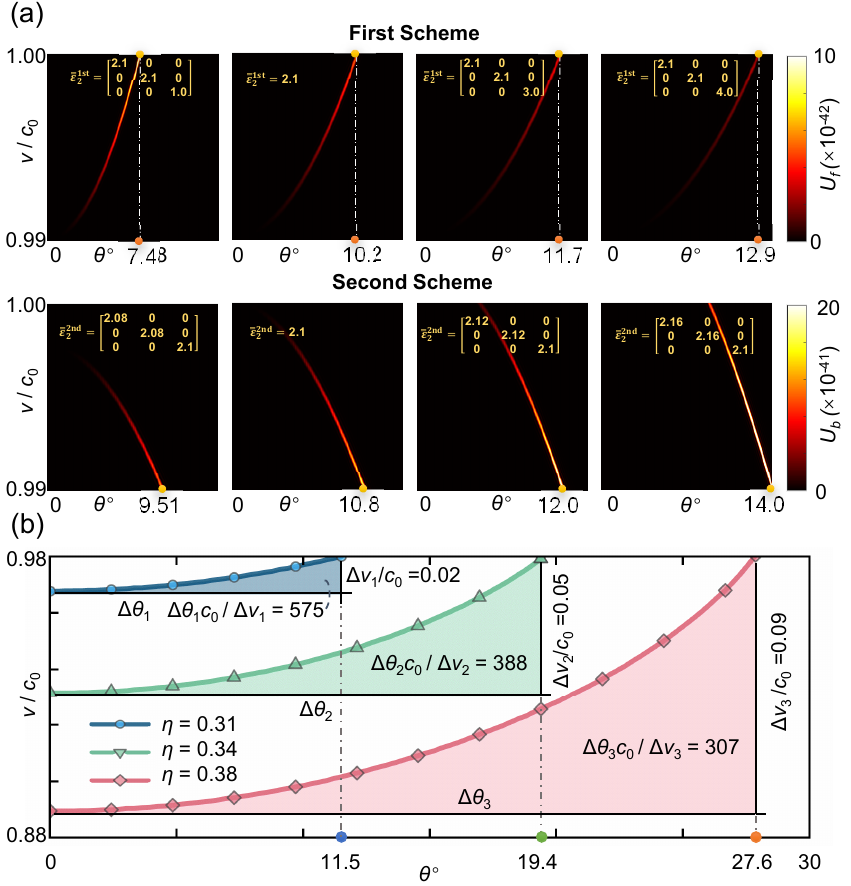}
    \caption{(a) Angular spectral energy density of the forward (backward) radiation $U_{f(b)}(\theta,0,\omega_0)$ for various configurations of PCs \cite{lin2018controlling}. The top panel shows the forward radiation patterns that corresponds to the first scheme, for which the thickness of the unit cell is $L_\text{unit}^\text{1st} = 1.0205\lambda_0$ with the filling ratio of $\varepsilon_1$ reading $\eta^\text{1st}=0.3$. The permittivity of the other uniaxial crystal reads $\tensor{\varepsilon}_2^\text{1st} = \text{diag}(\varepsilon_\text{2p}^\text{1st}, \varepsilon_\text{2p}^\text{1st}, \varepsilon_\text{2s}^\text{1st})$ with $\varepsilon_\text{2p}^\text{1st}=2.1$ and $\varepsilon_\text{2s}^\text{1st}=1.0$, 2.1, 3.0, 4.0, respectively. For the backward radiation patterns shown in the lower panel which corresponding to the second scheme, $L_\text{unit}^\text{2nd}=0.2792\lambda_0$ and $\eta^\text{2nd}=0.6$; and the permittivity of the uniaxial crystal reads $\tensor{\varepsilon}_2^\text{2nd} = \text{diag}(\varepsilon_\text{2p}^\text{2nd}, \varepsilon_\text{2p}^\text{2nd}, \varepsilon_\text{2s}^\text{2nd})$ with $\varepsilon_\text{2s}^\text{2nd}=2.1$ and $\varepsilon_\text{2p}^\text{2nd}=2.08$, 2.10, 2.12, 2.16, respectively. (b) The Cherenkov-like radiation angles with respect to the charge velocity for various filling ratios $\eta^\text{2nd}$ of 0.31, 0.34 and 0.38. Here, the PC configuration of $L_\text{unit} = 1.0205\lambda_0$, $\varepsilon_1=10.6$, and $\varepsilon_2=2.1$ is considered. In all the simulations, $\varepsilon_1 = 10.6$; the overall thickness of the multilayers (or PCs) is 2~\si{mm} and the working wavelength of interest is $\lambda_0 = 700~\si{nm}$.}
    \label{fig:figure03}
\end{figure}

We have validated the proposed generalized cascading scattering matrix method for multilayers constituting of uniaxial crystals by examining the PC-based Cherenkov-like radiation and comparing the obtained results with the literature \cite{lin2018controlling}. In the validation, the multilayered structure consists of an isotropic medium with $\varepsilon_1 = 10.6$ and a uniaxial crystal $\tensor{\varepsilon}_2$. For different filling ratios of the isotropic medium $\eta=d_1/L_\text{unit}$, where $L_\text{unit} = d_1 + d_2$ is the unit cell thickness with $d_i$ ($i=1,2$) being the thickness of the medium $\varepsilon_i$, different types of Cherenkov-like transition radiation could exhibit when the other material property $\varepsilon_2$ varies, as shown in \figref{fig:figure03}.

\figref{fig:figure03}(a) displays the angular spectral energy density of two types of Cherenkov-like transition radiation for different configurations of multilayers in unit cell thickness, filling ratios, and the cell material property. In the top panel, when the thickness of the unit cell is $L_\text{unit} = 1.0205\lambda_0$ and the filling ratio of $\varepsilon_1$ is $\eta=0.3$, as the on-axis permittivity of the other uniaxial crystal $\varepsilon_\text{2s}$ varies from 1.0 to 4.0 while the off-axis permittivity is fixed at $\varepsilon_\text{2p} = 2.1$, the detectable angle range of the Cherenkov-like transition radiation is increased from 7.48~\si{\degree} to 12.9~\si{\degree}. Specifically, when the uniaxial crystal $\varepsilon_2$ reduces to an isotropic medium, the calculated angular spectrum of forward radiation matches that presented in Ref. \cite{lin2018controlling} for PCs. Similarly, strong backward Cherenkov-like transition radiation can occur when $L_\text{unit}$, $\eta$ and $\varepsilon_2$ change, as depicted in the lower panel in \figref{fig:figure03}(b).

\figref{fig:figure03}(b) shows the so-called detection curve that corresponds to the peaks in \figref{fig:figure03}(a) for various filling ratios of PCs constituting of isotropic media. As the filling ratio $\eta$ increases, the velocity threshold of the Cherenkov-like transition radiation decreases and the observable angle grows. In contrast, the resolution, \emph{i.e.}, $\Delta \theta/(\Delta v/c_0)$, would decrease as $\eta$ is increasing; for example, it decreases from 575 at $\eta=0.31$ to 307 at $\eta=0.38$. When designing a PC-based Cherenkov radiation detector, a trade-off occurs between the particle velocity threshold (and range) and the detection accuracy.

\subsection{Cherenkov-like radiation in multilayers comprised of uniaxial crystals with tilting optical axes}
\begin{figure}[!ht]
    \centering
    \includegraphics{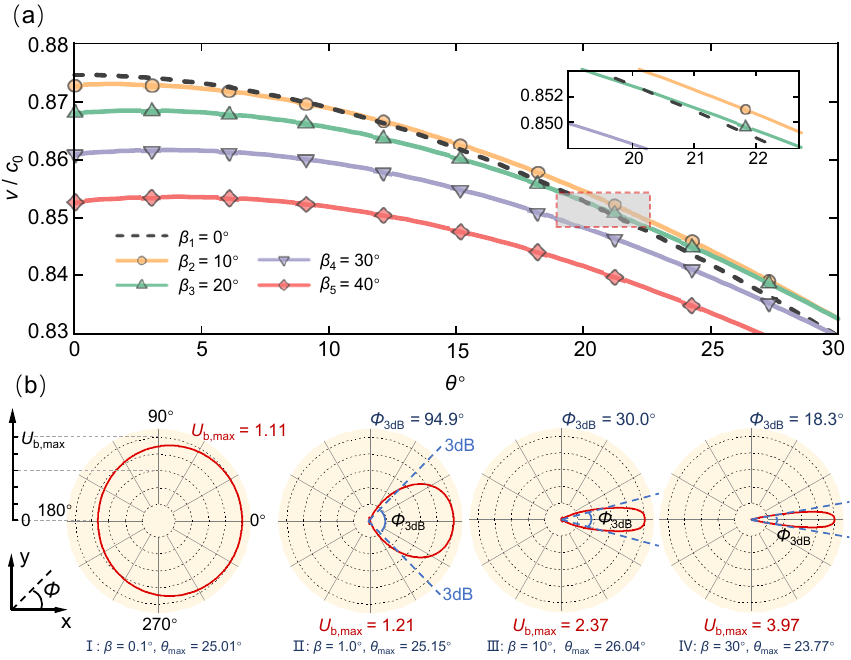}
    \caption{(a) The peaks of angular spectral energy density of the Cherenkov-like resonant transition radiation in uniaxial-crystal-based PCs, for which the optical axes is tilted by an angle $\beta$ of 10~\si{\degree} (circles), 20~\si{\degree} (up-pointing triangles), 30~\si{\degree} (down-pointing triangles) and 40~\si{\degree} (diamonds), respectively, with respect to $z$-axis (direction of the electron trajectory). As a reference, the results for uniaxial-crystal-based PC without titled is given in dashed curves ($\beta = 0~\si{\degree}$). For the multilayered PC, the thickness of unit cell is $L_\text{unit}=1.02\lambda_0$ with the working wavelength $\lambda_0=590~\si{nm}$; and the number of unit cells is $N_c = 2000$, yielding a overall thickness 1.2036~\si{mm}. The filling ratio of GaP is $\eta = 0.7$. (b) The corresponding radiation patterns in the horizontal plane parallel to the interface. Here, $\theta$ denotes the azimuth angle of the Cherenkov-like radiation where the maximum spectral density occurs. Note that, all the radiation patterns $U_f(\theta=\theta_\text{max}^\beta, \phi, \omega_0)$ are normalized to the isotropic spectral density $U_f(\theta_\text{max}^{\beta = 0}=25.01~\si{\degree}, 0, \omega_0)$. In addition, the half-power beamwidth (3dB angle) $\Phi_\text{3dB}$ is calculated. The electron velocity is $v = 0.8420 c_0$ and $\omega_0 = 2\pi c_0/\lambda_0$. }
    \label{fig:figure04}
\end{figure}

The Cherenkov-like radiation pattern depends on the proper configurations of the geometry and material properties; thus, by adding a new degree of freedom for the uniaxial-crystal-based PC design, \emph{i.e.}, the angle of the optical axis, it is possible to realize new types of Cherenkov-like radiation. \figref{fig:figure04} illustrates the Cherenkov-like resonant transition radiation in PCs that are made of alternating layers of a high-dielectric constant (isotropic) material and a uniaxial crystal with tilting optical axis. Actual dielectrics GaP and $\text{CaCO}_3$ are used as the isotropic medium and uniaxial crystal, with relative permittivities $\varepsilon_1 = 10.6$, $\varepsilon_\text{2p} =2.749 $ and $\varepsilon_\text{2s}=2.208$ at the wavelength of interest $\lambda_0=590~\si{nm}$. 

\figref{fig:figure04}(a) plot the detection curves corresponding to the bright line in the angular spectral energy density when the optical axes of $\text{CaCO}_3$ for the multilayers are titled by an angle $\beta$ with respect to $z$-axis. Now, the transform $\tensor{\varepsilon}_2 = \Lambda^{-1}(0,\beta,0) \text{diag}(\varepsilon_\text{2p}, \varepsilon_\text{2p}, \varepsilon_\text{2s}) \Lambda(0,\beta,0)$ is applied when calculating the optical properties of the GaP-$\text{CaCO}_3$-made PCs, where $\Lambda(0,\beta,0)$ denotes the coordinate rotation Euler matrix. It is evident that the maximum detectable velocity $v_\text{max}$ decreases as the inclination angle $\beta$ decreases. Moreover, as the tilting angle increases, the detection curve becomes flatter, implying that greater velocity resolution can be achieved.

As shown in \figref{fig:figure04}(b), when the optical axis of the uniaxial crystal is tilted, the PCs no longer have rotational symmetry, and directional radiation would appear in the horizontal planes normal to the electron trajectory. Intriguingly, with only a minor tilting angle $\beta$, the radiation acquires strong directionality. As the tilting angle increases, the corresponding half-power beamwidth (3dB angle) $\Phi_\text{3dB}$ decreases, indicating nearly improved orientation in the $x$ direction. Meanwhile, the maximum detectable energy density $U_{f,\text{max}}$ is significantly greater than that without tilting, indicating that Cherenkov-like radiation may be easier to observe in such a structure. 
Since the pitch and roll angles are zero for the Euler matrix $\Lambda(0,\beta,0)$ in the simulation, the radiation patterns $U_f(\theta_\text{max}^\beta, \phi, \omega_0)$ are symmetric with respect to $x$-axis (\emph{viz.}, $\phi = 0$).

\subsection{Tandem-network-assisted design of angular spectrum}
By breaking the rotation symmetry of materials with respect to the direction of the charge trajectory, it may be possible to excert comprehensive control over the Cherenkov-like transition radiation in multilayers. However, it would add complexity to the design. Moreover, as mentioned previously, analyzing and designing multilayered structures \emph{per se} such as PCs can be difficult. As shown in \figref{fig:figure03} and \figref{fig:figure04}, similar electromagnetic properties exist for various configurations of geometry (\emph{e.g.}, unit cell thickness, filling ratio, periodicity, \emph{etc.}) and materials (isotropic or anisotropic). Deep neural networks provide a promising solution to this problem in this paper. The parameters of the Tandem neural network are summarized in \tabref{tab:table01}.
\begin{table}[!ht]
  \centering
  \begin{ThreePartTable}
  \caption{\bf Detailed information for the Tandem networks.}  \label{tab:table01}
  \begin{tabularx}{3.4in}{ccc} \toprule
                  & Forward net & Inverse net \\ \midrule
    NN architecture  & $3/512(4)/20$ \tnote{a} & $20/512/1024(3)/512/3$ \tnote{b}\\
    Activation functions & Leaky ReLu & Leaky ReLu \\
    Optimizer & Adam & Adam \\
    Learning rate & $10^{-5}$ & $10^{-5}$ \\
    Loss function & MSE \tnote{c} & MSE \& Smooth L1 \tnote{d}\\
  \bottomrule
\end{tabularx}
  \begin{tablenotes} \footnotesize
    \item[a] $512(4)$ denotes four hidden layers, each with $512$ neurons.
    \item[b] $512/1024(3)/512$ denotes two hidden layers, each with 512 neurons and three hidden layers, each with $1024$ neurons.
    \item[c] Mean square error: $\text{MSE}(\bm{y},\bm{y}') = \mathbb{E}(\bm{y}-\bm{y}')$, where $\bm{y}$ is the actual optical response (input) in the training dataset obtained according to \algref{alg:AnalyticalGCSMM}, and $\bm{y}'$ is the trained optical response derived from the neural networks (output).
    \item[d] $\text{MSE}(\bm{y},\bm{y}')+\kappa \text{Smooth L1}(\bm{x},\bm{x}')$, where $\kappa = 0.01$ in our model. Interested readers can refer to \cite{Girshick2015ICCV} for the detailed information of Smooth L1. Here, $\bm{y}$ is the expected optical response (input) and $\bm{y}'$ is the designed optical response derived from the inverse net; $\bm{x}$ and $\bm{x}'$ are the structure parameters corresponding to $\bm{y}$ and $\bm{y}'$, respectively.
  \end{tablenotes}
  \end{ThreePartTable}
\end{table}

The objective of this paper is to design Cherenkov-like radiation based on PC, which constitutes of alternating layers of an isotropic medium and a uniaxial crystal with tilting optical axis. As shown in \figref{fig:figure02}(b), the input $\bm{y}$ is now the detection curve over a range of angle $\theta$, where only the data between $\theta \in [18~\si{\degree}, 24~\si{\degree}]$ have been adopted. Here, the evenly sampled $M = 20$ data are represented by $(x_i = \theta_i, y_i = v_i/c_0)$ ($i=1, ...,M$). The output of the overall Tandem network $\bm{y}'$ has the same physical meanings as the input $\bm{y}$, which is the predicted (calculated) response for the designed parameter $\bm{x}'$ (\emph{i.e.}, the mid-products of the Tandem network). The designing parameters $\bm{x}'$ to be determined include the filling ratio $\eta$, the thickness of the unit cell $L_{\text{unit}}$, and tilting angle $\beta$ of the uniaxial crystal with respect to $z$-axis, \emph{i.e.}, $\bm{x}'= \{ \eta, L_{\text{unit}}, \beta \}$. Note that $\bm{x}'$ is the output of the inverse network and the input to the forward network. During the training, $\{\bm{x},\bm{y}\}$ is referred to as the training element; however, $\bm{x}$ becomes the designed parameters $\bm{x}'$ during the design process.

The deep learning framework \emph{PyTorch} \cite{paszke2019pytorch} is adopted. The training dataset was produced by uniformly sampling parameter tensor $\bm{x}=[\eta,L_{\text{unit}},\beta]$ in the parameter space with $\eta \in [0.68, 0.71]$, $L_\text{unit} \in [1.02, 1.06]$ and $\beta \in [0~\si{\degree}, 60~\si{\degree}]$ and calculating the corresponding response tensor $\bm{y}$ using the proposed generalized scattering matrix method, as shown in \algref{alg:AnalyticalGCSMM}. For the to-be-designed PC, the actual dielectrics GaP and $\text{CaCO}_3$ are utilized, along with $N_c = 4000$ unit cells. 8000 data sets are divided into training, validation and test sets in a ratio of 3:1:1  during the training process for both the forward network and the Tandem network. In \tabref{tab:table01}, the loss function and other meta-parameters for the Tandem network are listed. 

\begin{figure}[!ht]
    \centering
    \includegraphics{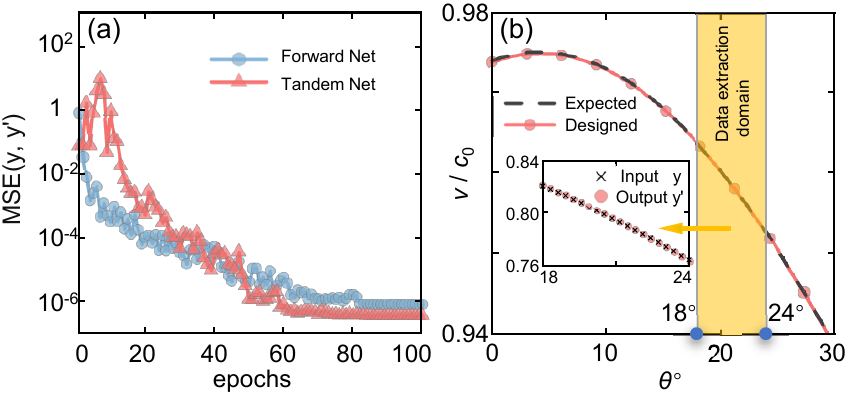}
    \caption{(a) The evolution of validation losses per epoch during the train process for the forward neural network (blue) and the overall Tandem network (red). (b) Comparison between the designed angular spectrum (dots) and the desired one (dashed curves). The inset depicts the portion of data in the shaded region that is used to train the neural network.}
    \label{fig:figure05}
\end{figure}

\figref{fig:figure05}(a) shows the validation error with respect to epochs, which is defined as the mean square error (MSE) between the expected and trained responses, \emph{i.e.}, $\text{error} = \text{MSE}(\bm{y},\bm{y}')$. For the forward network, the error for the validation dataset quickly converges to the order of $10^{-7}$ at after about 80 training epochs, indicating that it is capable of predicting the angular spectrum (response, $\bm y$) according to the geometry and material parameters ($\bm x$) rather than the conventional electromagnetic simulation methods. Moreover, the validation error converges for the Tandem networks, suggesting that the one-to-many design problem can also be effectively addressed. 

\figref{fig:figure05}(b) compares the designed angular spectrum to the desired response. We have obtained $\bm{x}'=(0.7064, 1.0577, 46.5060~\si{\degree})$ from the inverse network within a subset of parameter space used to train the Tandem network (see the inset in \figref{fig:figure05}); meanwhile, it yields nearly the same spectrum (circle) as the desired one (dashed curves) over the entire angle ranges of interest, which is calculated according to \algref{alg:AnalyticalGCSMM}. We notice that by only learning the local features, NNs can imitate the physical information hidden beneath a large dataset. In addition, the physical information contained in the model (\emph{i.e.}, the pre-trained forward network) provides crucial foundations for the final convergence of the entire network, which are unavailable by traditional data-driven neural networks.

\section{Conclusion}
In summary, we have proposed a general cascading scattering matrix method for calculating the electromagnetic properties induced by a moving charge traversing planar multilayered structures, which can be comprised of uniaxial crystals with tilted optical axes. In the external vacuum of photonic crystals with the proper configurations, Cherenkov-like resonant transition radiation can be observed, whose energy distribution can be used to detect particles' velocity. Incorporating uniaxial crystals and appropriately adjusting their optical axes enable the external radiation to be strongly directional in the plane normal to the electron trajectory, thereby introducing a new degree of freedom when designing PC-based Cherenkov detectors. Furthermore, PCs with a tilting optical axis can attain a higher resolution of velocity. In addition, with the help of Tandem networks, we are able to design PCs according to the desired detection curve, which demonstrates a significant improvement over conventional inverse neural networks by considering local sampling of training data and incorporating errors for both the design parameters and the response. We anticipate that the proposed Swiss army knives will benefit other PC applications in nanophotonics community.

\begin{backmatter}
\bmsection{Funding} National Natural Science Foundation of China (51977165).
\smallskip

\bmsection{Disclosures} The authors declare no conflicts of interest.
\smallskip

\bmsection{Data availability} Data underlying the results presented in this paper may be obtained from the authors upon reasonable request.
\smallskip


\end{backmatter}

\bibliography{main}

\end{document}